\documentclass{article}%
\usepackage{amsfonts}
\usepackage{amsmath}%
\usepackage{amssymb}%
\usepackage{graphicx}
\newtheorem{theorem}{Theorem}

\newtheorem{corollary}[theorem]{Corollary}

\newtheorem{definition}[theorem]{Definition}

\newtheorem{lemma}[theorem]{Lemma}

\newtheorem{proposition}[theorem]{Proposition}
\newtheorem{remark}[theorem]{Remark}

\newenvironment{proof}[1][Proof]{\textbf{#1.} }{\ \rule{0.5em}{0.5em}}
\begin{document}

\title{Geometric reduction of Hamiltonian systems}
\author{Krzysztof Marciniak\thanks{Partially supported by The Swedish Research Council
grant No. 624-2003-607}\\Department of Science and Technology \\Campus Norrk\"{o}ping, Link\"{o}ping University\\601-74 Norrk\"{o}ping, Sweden
\and Maciej B\l aszak\thanks{Partially supported by The Swedish Institute
scholarship No. 03824/2003 and by KBN Research Project 1 P03B 111 27}\\Institute of Physics, A. Mickiewicz University\\Umultowska 85, 61-614 Pozna\'{n}, Poland}
\maketitle

\begin{abstract}
Given a foliation $\mathcal{S}$ of a manifold $\mathcal{M}$, a distribution
$\mathcal{Z}$ in $\mathcal{M}$ transveral to $\mathcal{S}$ and a Poisson
bivector $\Pi$ on $\mathcal{M}$ we present a geometric method of reducing this
operator on the foliation $\mathcal{S}$ along the distribution $\mathcal{Z}$.
It encompasses the classical ideas of Dirac (Dirac reduction) and more modern
theory of J. Marsden and T. Ratiu, but our method leads to formulas that allow
for an explicit calculation of the reduced Poisson bracket. Moreover, we
analyse the reduction of Hamiltonian systems corresponding to the bivector
$\Pi$.

\end{abstract}


\section{Introduction}

The reduction theory of dynamical systems consist of two branches: the first
branch deals with constrained Lagrangian systems, the second with constrained
Hamiltonian systems. In the Lagrangian approach one consider separately the
case of holonomic constraints, i.e. the constraints which may depend on
velocities, but only in such a way that the equations of constraints can be
integrated to eliminate velocities, and the non-holonomic case. In many of
these papers one first considers the Lagrangian formulation and then passess
to the corresponding Hamitlonian formulation (see for examlple \cite{Van der
Schaft}). The reduction theory in the Hamiltonian context has been initiated
by P.A.M. Dirac, who in his famous paper \cite{Dirac} described a method of
reducing a given Poisson bracket onto a submanifold given by some constraints
$\varphi$ provided they were of "second class". In this approach the classical
notion of holonomic constraints is usually not introduced as in this context
there is no obvious division of variables between "position" and "velocity"
(or "momenta"). Recently, there has been much interest to extend the theory
onto the case of generalized Hamiltonian systems (see for example \cite{blan}).

The ideas of Dirac were developed in many papers, among others in
\cite{s1}-\cite{fla} (see also the literature quoted there). A geometric
meaning of this reduction procedure has been investigated in
\cite{MarsdenRatiu} and in \cite{Diracrevisited}.

In this paper we develop the ideas of \cite{Dirac} and \cite{MarsdenRatiu} and
present a constructive, computable method of reducing (locally) a given
Poisson operator $\Pi$ to any regular submanifold $\mathcal{S}$. The idea of
the method is to choose a distribution $\mathcal{Z}$ (not necessarily
integrable) that is i) transversal to the foliation $\mathcal{S}$ ii) at any
$x\in\mathcal{M}$ it completes $T_{x}\mathcal{S}$ to $T_{x}\mathcal{M}$ iii)
it makes the operator $\Pi$ $\mathcal{Z}$-invariant (see definitions below)
and then to deform the Poisson operator $\Pi$ to a new Poisson operator
$\Pi_{D}$ such that its image will be tangent to the submanifold $\mathcal{S}%
$. This new operator $\Pi_{D}$ will be always Poisson (and so its natural
restriction to $\mathcal{S}$ will be Poisson). In consequence, we obtain a
method of reducing a Hamiltonian system on $\mathcal{M}$ to a Hamiltonian
system on every leaf $\mathcal{S}_{\nu}$ of the foliation $\mathcal{S}$.\ This
reduced system strongly depends on the choice of the distribution
$\mathcal{Z}$. As a special case we obtain the classical Dirac reduction of
Hamiltonian system. All our considerations will be local in the sense that our
manifold $\mathcal{M}$ is perhaps only an open submanifold of a larger
manifold. Our construction is equivalent to the reduction method proposed by
Marsden and Ratiu in \cite{MarsdenRatiu}. However, our approach has
advantages: it can be performed simultaneously on any leaf $\mathcal{S}_{\nu}$
of the foliation $\mathcal{S}$, it is constructive (the approach of Marsden
and Ratiu requires calculations of prolongations of $\mathcal{Z}$-invariant
functions and as such is difficult to perform in practice) and it is
formulated in the language of Poisson bivectors rather than Poisson brackets.

We want to stress that this method does not require the submanifold to be
given by a holonomic constraints on some configuration space. In fact, we do
not require our manifold to be a cotangent bundle to any configurational
manifold at all, but of course our construction covers this special case as
well. For example, it covers the case discussed in \cite{Van der Schaft},
where the auhtors obtain the Poisson operator on the constrained submanifold
only in the case of holonomic constraints - because they simply restrict their
Poisson operator $\Pi$ onto the constrained submanifold, and such restriction
usually (apart form the holonomic case) destroys the poissonity of the operator.

We have to notice that some particular versions of the proposed scheme
appeared recently in \cite{degiovanni} and \cite{falquipedroni}. In
\cite{Marle} the author applied the same setting as above but with no
conclusive formulas for computing the actual deformed Poisson bracket $\Pi
_{D}$ and since he did not use the notion of $\mathcal{Z}$-invariance, his
deformed operator $\Pi_{D}$, although formally identical with our
construction, was not always Poisson (it has been called there a
pseudo-Poisson operator).

Some basic steps of our construction have been presented in our previous paper
\cite{dualpp}. This paper, however, was mainly devoted to completion of the
above picture by its "dual" part by developing a theory of a Marsden-Ratiu
type reduction of presymplectic 2-forms $\Omega$ that are (in a sense) dual to
a given Poisson operator $\Pi$. The picture presented here is much clearer and
moreover it is paramietrization-free in the sense that we prove the main
results without neccessity of dicsussing some particular functions defining
our foliation $\mathcal{S}$. Moreover, in the paper \cite{dualpp} we focused
on Dirac case while this paper has a general character. In the end, in this
paper we also consider the related reduction of Hamiltonian dynamics.

\section{Geometric reduction of Poisson bivectors}

Let us consider a smooth manifold $\mathcal{M}$ of arbitrary (finite)
dimension $n$ and a foliation $\mathcal{S}$ of $\mathcal{M}$ consisting of the
leaves $\mathcal{S}_{\nu}$ parametrized by $\nu\in\mathbf{R}^{k}$ (so that
$k\in\mathbf{N}$ is the codimension of every leaf $\mathcal{S}_{\nu}$).
Consider also a regular distribution $\mathcal{Z}$ on $\mathcal{M} $ (that is
a smooth collection of the spaces $\mathcal{Z}_{x}\subset T_{x}\mathcal{M}$
where $\nu$ is such that $x\in\mathcal{S}_{\nu}$) such that it completes every
$T_{x}\mathcal{S}$ to $T_{x}\mathcal{M}$ in the sense that%
\begin{equation}
T_{x}\mathcal{M=}T_{x}\mathcal{S}_{\nu}\oplus\mathcal{Z}_{x}\label{split}%
\end{equation}
for every $x$ in$\mathcal{M}$. Here and in what follows $\oplus$ denotes the
direct sum of vector spaces. It means that every vector field $X$ on the
manifold $\mathcal{M}$ has a unique decomposition $X=X_{\parallel}+X_{\perp}$
such that for every $x$ in $\mathcal{M}$ the vector $\left(  X_{\parallel
}\right)  _{x}\in T_{x}\mathcal{S}$ ($X_{\parallel}$ is tangent to the leaves
of the foliation $\mathcal{S}$) while $\left(  X_{\perp}\right)  _{x}%
\in\mathcal{Z}_{x}$ ($X_{\perp}$ is contained in the distribution
$\mathcal{Z}$). The splitting (\ref{split}) induces the following splitting of
the corresponding dual space $T_{x}^{\ast}\mathcal{M}$:%

\begin{equation}
T_{x}^{\ast}\mathcal{M=}T_{x}^{\ast}\mathcal{S}_{\nu}\oplus\mathcal{Z}%
_{x}^{\ast}\label{dualsplit}%
\end{equation}
where $T_{x}^{\ast}\mathcal{S}_{\nu}$ is the annihilator of $\mathcal{Z}_{x}$
while the space $\mathcal{Z}_{x}^{\ast}$ is the annihilator of $T_{x}%
\mathcal{S}_{\nu}$. Thus, any one-form $\alpha$ on $\mathcal{M}$ has a unique
decomposition $\alpha=\alpha_{\parallel}+\alpha_{\perp}$ such that $\left(
\alpha_{\parallel}\right)  _{x}\in T_{x}^{\ast}\mathcal{S}$ ($\alpha
_{\parallel}$ annihilates the vectors from $\mathcal{Z}$) while $\left(
\alpha_{\perp}\right)  _{x}\in\mathcal{Z}_{x}^{\ast}$ ($\alpha_{\perp}$
annihilates the vectors tangent to the foliation $S$ ). We will call
$X_{\parallel}$ and $\alpha_{\parallel}$ as projections of $X$ and $\alpha$
(respectively) on the foliation $\mathcal{S}$. Abusing notation a bit we will
write that $X\subset T\mathcal{S}$ if $X=X_{\parallel}$, $X\subset\mathcal{Z}$
if $X=X_{\perp}$ and similarly for one forms: $\alpha\subset T^{\ast
}\mathcal{S}$ if $\alpha=\alpha_{\parallel}$, $\alpha\subset\mathcal{Z}^{\ast
}$ if $\alpha=\alpha_{\perp}$.

Let us now suppose that our manifold $\mathcal{M}$ is equipped with a Poisson
bivector $\Pi$ (i.e. a bivector with vanishing Schouten bracket, see
\cite{Lichnerowicz}). This operator induces the Poisson bracket $\left\{
F,G\right\}  _{\Pi}=\left\langle dF,\Pi dG\right\rangle $ on the algebra of
smooth functions on $\mathcal{M}$, where $\left\langle \cdot,\cdot
\right\rangle $ is the dual map between $T^{\ast}\mathcal{M}$ and
$T\mathcal{M}$. A smooth real-valued function $F$ on $\mathcal{M}$ \ is called
$\mathcal{Z}$-invariant if the Lie derivative $L_{Z}F=0$ for any vector field
$Z\subset\mathcal{Z}$. We will now adopt the following definition.

\begin{definition}
The operator $\Pi$ is said to be $\mathcal{Z}$-invariant if $L_{Z}\left\{
F,G\right\}  _{\Pi}=0$ for any pair of $\mathcal{Z}$-invariant functions $F$
and $G$ and every vector field $Z\subset\mathcal{Z}$.
\end{definition}

Notice that our definition does not necessarily mean that $L_{Z}\Pi=0$ for all
vector fields $Z\subset\mathcal{Z}$, as for any pair $F,G$ of $\mathcal{Z}%
$-invariant functions the condition $L_{Z}\left\{  F,G\right\}  _{\Pi}=0$
means only that the function $\left\langle dF,\left(  L_{Z}\Pi\right)
dG\right\rangle $ vanishes. Thus, $\Pi$ does not have to be an invariant of
the distribution $\mathcal{Z}$ to be $\mathcal{Z}$-invariant in our meaning.
Notice also, that the above definition is equivalent to the statement that for
any pair $\alpha,\beta$ $\subset$ $T^{\ast}\mathcal{S}$ we have $\left\langle
\alpha,\left(  L_{Z}\Pi\right)  \beta\right\rangle =0$ (since if $F$ is
$\mathcal{Z}$-invariant then $dF\subset T^{\ast}\mathcal{S}$).

Suppose for the moment that the distribution $\mathcal{Z}$ is spanned by $k$
vector fields $Z_{i}$. We say, that the operator $\Pi$ is $\emph{Vaisman}$
\cite{Vaisman} with respect to $\mathcal{Z}$ if for every vector field $Z_{i}
$ there exists vector fields $W_{ij}$, $j=1,\ldots k$ such that%
\begin{equation}
L_{Z_{i}}\Pi=%
{\textstyle\sum\limits_{j=1}^{k}}
W_{ij}\wedge Z_{j}.\label{vais}%
\end{equation}
It is easy to see that this definition does not depend on the choice of basis
in $\mathcal{Z}$ (although the vector fields $W_{ij}$ obviously do). If the
operator $\Pi$ is Vaisman with respect to $\mathcal{Z}$, then it is also
$\mathcal{Z}$-invariant, as then for any two one-forms $\alpha,\beta\subset
T^{\ast}\mathcal{S}$
\[
\left\langle \alpha,\left(  L_{Z}\Pi\right)  \beta\right\rangle =%
{\textstyle\sum\limits_{j=1}^{k}}
\left\langle \alpha,\left(  W_{ij}\wedge Z_{j}\right)  \beta\right\rangle =0,
\]
since $\alpha$ and $\beta$ annihilate all the vector fields $Z_{i}$. The
converse statement is however not true in general.

Let us now consider a Poisson operator $\Pi$ on $\mathcal{M}$ and define the
following bivector:%
\begin{equation}
\Pi_{D}\left(  \alpha,\beta\right)  =\Pi(\alpha_{||},\beta_{||})\text{ \ for
any pair }\alpha,\beta\text{ of one-forms.}\label{defpi}%
\end{equation}
We will often call the bivector $\Pi_{D}$ a \emph{deformation} of $\Pi$. This
bivector occurred for example in \cite{degiovanni} in a more restrictive
context. Observe that it always exists and that it is uniquely defined once
the foliation $\mathcal{S}$ and the distribution $\mathcal{Z}$ are given (it
is thus a purely geometric construction). It has an important property: its
image lies always in $T\mathcal{S}$.

\begin{lemma}
$\Pi_{D}(\alpha)\subset T\mathcal{S}$ for any one-form $\alpha$ on
$\mathcal{M}$, i.e. the image of $\Pi_{D\text{ }}$is tangent to the foliation
$\mathcal{S}$.
\end{lemma}

\begin{proof}
We have to show that $\left\langle \beta,\Pi_{D}\alpha\right\rangle =0$ for
any $\beta\subset Z^{\ast}$. But%
\[
\left\langle \beta,\Pi_{D}\alpha\right\rangle =\Pi_{D}\left(  \beta
,\alpha\right)  =\Pi\left(  \beta_{\parallel},\alpha_{\parallel}\right)  =0
\]
since $\beta_{\parallel}=0$ for every $\beta\subset Z^{\ast}$.
\end{proof}

Thus, the deformed bivector $\Pi_{D\text{ }}$ has its image in $T\mathcal{S} $
and if we consider it as mapping from one-forms to vector fields on
$\mathcal{M}$ then it can be naturally restricted to a bivector $\pi_{R_{\nu}%
}$ on every leaf $\mathcal{S}_{\nu}$ of $\mathcal{S}$ by simply restricting
its domain to $\mathcal{S}_{\nu}:$%
\[
\pi_{R_{\nu}}=\left.  \Pi_{D}\right\vert _{\mathcal{S}_{\nu}}.
\]
Moreover, it induces a new bracket for functions on $\mathcal{M}$:%
\[
\left\{  F,G\right\}  _{\Pi_{D}}=\Pi_{D}\left(  dF,dG\right)  =\Pi(\left(
dF\right)  _{||},\left(  dG\right)  _{||}).
\]
Of course, the bivector $\Pi_{D\text{ }}$(and thus even $\pi_{R_{\nu}}$) in
general does not have to be Poisson. However, it turns out that if $\Pi$ is
$\mathcal{Z}$-invariant then $\Pi_{D}$ (and thus every $\pi_{R_{\nu}}$) is Poisson.

\begin{theorem}
If $\Pi$ is $\mathcal{Z}$-invariant then $\Pi_{D}$ given by (\ref{defpi}) is Poisson.
\end{theorem}

\begin{proof}
We will prove that the bracket $\left\{  \cdot,\cdot\right\}  _{\Pi_{D}}$ is
Poisson. Obviously, this bracket is antisymmetric and satisfies the Leibniz
property. It remains to show, that it also satisfies the Jacobi identity, that
is $\left\{  \left\{  F,G\right\}  _{\Pi_{D}},H\right\}  _{\Pi_{D}}+$ cycl.
$=0$, for any functions $F,G,H$. Using the definition of $\Pi_{D}$, this
condition can be written as
\begin{equation}
\left\langle \left(  d\left\langle \left(  dF\right)  _{\parallel},\Pi\left(
dG\right)  _{\parallel}\right\rangle \right)  _{\parallel},\Pi\left(
dH\right)  _{\parallel}\right\rangle +\text{cycl.}=0.\label{condition}%
\end{equation}
However, for any vector field $Z\subset\mathcal{Z}$ we have%
\[
\left\langle d\left\langle \left(  dF\right)  _{\parallel},\Pi\left(
dG\right)  _{\parallel}\right\rangle ,Z\right\rangle =Z\left(  \left\langle
\left(  dF\right)  _{\parallel},\Pi\left(  dG\right)  _{\parallel
}\right\rangle \right)  =L_{Z}\left\langle \left(  dF\right)  _{\parallel}%
,\Pi\left(  dG\right)  _{\parallel}\right\rangle =0
\]
due to the assumed $\mathcal{Z}$-invariance of $\Pi$. This means that
$d\left\langle \left(  dF\right)  _{\parallel},\Pi\left(  dG\right)
_{\parallel}\right\rangle \subset T^{\ast}\mathcal{S}$, so that
\[
\left(  d\left\langle \left(  dF\right)  _{\parallel},\Pi\left(  dG\right)
_{\parallel}\right\rangle \right)  _{\parallel}=d\left\langle \left(
dF\right)  _{\parallel},\Pi\left(  dG\right)  _{\parallel}\right\rangle ,
\]
and thus condition (\ref{condition}) turns out to be the Jacobi identity for
$\Pi$, which is satisfied since $\Pi$ is Poisson.
\end{proof}

Thus, given a foliation $\mathcal{S}$ on $\mathcal{M}$ and a transversal
distribution $\mathcal{Z}$ on $\mathcal{M}$ (such that (\ref{split}) is
satisfied) we can reduce any Poisson bivector $\Pi$ that is $\mathcal{Z}%
$-invariant to a Poisson bivector $\pi_{R_{\nu}}$ on the leaf $\mathcal{S}%
_{\nu}$ of $\mathcal{S}$ by deforming $\Pi$ to $\Pi_{D}$ and then by
restricting $\Pi_{D}$ to $\mathcal{S}_{\nu}$. This construction yields the
same operator $\pi_{R_{\nu}}$ as in the approach of Marsden and Ratiu
\cite{MarsdenRatiu}. We will however show how this construction can be easily
realized in practice.

\begin{remark}
In case when the foliation $\mathcal{S}$ coincides with the symplectic
foliation of $\Pi$ we have of course that $\Pi_{D}=\Pi\,$, since in this case
$\Pi((dF)_{\perp})=0$ for any function $F$ and so $\left\{  F,G\right\}
_{\Pi_{D}}=\Pi(\left(  dF\right)  _{||},\left(  dG\right)  _{||})=\Pi\left(
dF,dG\right)  =\left\{  F,G\right\}  _{\Pi}$. In this case $\pi_{R_{\nu}}$ is
the standard reduction of $\Pi$ on its symplectic leaf $\mathcal{S}_{\nu}$.
\end{remark}

Let us now consider some special cases of our general situation. We firstly
observe that the annihilator $\mathcal{Z}^{\ast}$ of $T\mathcal{S}$ is defined
as soon as the foliation $\mathcal{S}$ is determined (we do not need to
specify a particular $\mathcal{Z}$ in order to define $\mathcal{Z}^{\ast} $).

\begin{definition}
The distribution $\mathcal{D}=\Pi\left(  \mathcal{Z}^{\ast}\right)  $ (so that
$\mathcal{D}_{x}=\Pi\left(  \mathcal{Z}_{x}^{\ast}\right)  $ ) is called a
Dirac distribution associated with the foliation $\mathcal{S}$.
\end{definition}

Thus, the distribution $\mathcal{D}$ is determined by $\mathcal{S}$ and by
$\Pi$. A priori, two limit cases are possible here. If $T\mathcal{M=D}\oplus
T\mathcal{S}$ we say that we are in the \emph{Dirac case}. If \ $\mathcal{D}%
\subset T\mathcal{S}$ we say that we are in the \emph{tangent case}. In the
Dirac case we have a canonical choice of $\mathcal{Z}$: we can choose
$\mathcal{Z}=\mathcal{D}$ (in this case $\Pi$ is automatically $\mathcal{Z}%
$-invariant, since $\mathcal{Z}$ is spanned by the vector fields Hamiltonian
with respect to $\Pi$). Nevertheless, we can also choose some other
distribution $\mathcal{Z}\neq\mathcal{D}$. In the tangent case we have no
canonical choice of $\mathcal{Z}$ and we are free to find a distribution
$\mathcal{Z}$ that makes $\Pi$ $\mathcal{Z}$-invariant. Anyway, in both cases
we have many non-equivalent deformations $\Pi_{D}$ (and thus projections
$\pi_{R_{\nu}}$). Generically, the distribution $\mathcal{D}$ will not be
tangent to $\mathcal{S}$, but it will not suffice to span $T\mathcal{M}$
together with $T\mathcal{S}$.

Let us now suppose that the foliation $\mathcal{S}$ of $\mathcal{M}$ is
parametrized by the set of $k$ functionally independent real valued functions
$\varphi_{i}(x)$ so that its leaves have the form $\mathcal{S}_{\nu}=\left\{
x\in\mathcal{M}:\varphi_{i}(x)=\nu_{i}\text{, }\nu_{i}\in\mathbf{R}\text{,
}i=1,\ldots,k\,\right\}  $ where $k$ is - as above - the codimension of the
foliation. We will show how the above considerations can be written in the
parametrization $\left\{  \varphi_{i}\right\}  $. The one-forms $d\varphi_{i}$
constitute a basis in $\mathcal{Z}^{\ast}$. Then, the Dirac distribution
$\mathcal{D}$ is spanned by $k$ (possibly dependent) Hamiltonian vector fields
$X_{i}=\Pi d\varphi_{i}$. Let us denote a basis of $\mathcal{Z}$ dual to the
basis $\left\{  d\varphi_{i}\right\}  $ in $\mathcal{Z}^{\ast}$ by $Z_{i}$, so
that $Z_{i}(\varphi_{j})=\delta_{ij}$. Our projections $X_{\parallel}$ and
$\alpha_{\parallel}$ are then given by
\[
X_{||}=X-%
{\textstyle\sum\limits_{i=1}^{k}}
X(\varphi_{i})Z_{i},
\]
(obviously $X_{||}(\varphi_{i})=0$ for all $i$ so that indeed this vector
field is tangent to the leaves of $\mathcal{S}$) and by%
\[
\alpha_{||}=\alpha-%
{\textstyle\sum\limits_{i=1}^{k}}
\alpha(Z_{i})d\varphi_{i}%
\]
(and obviously $\alpha_{||}(Z_{i})=0$ for all $i$). Thus
\begin{align*}
\Pi(\alpha_{||},\beta_{||})  & =\Pi\left(  \alpha-%
{\textstyle\sum\limits_{i=1}^{k}}
\alpha(Z_{i})d\varphi_{i},\beta-%
{\textstyle\sum\limits_{j=1}^{k}}
\beta(Z_{j})d\varphi_{j}\right)  =\\
& =\Pi(\alpha,\beta)-%
{\textstyle\sum\limits_{j=1}^{k}}
\beta(Z_{j})\Pi(\alpha,d\varphi_{j})-%
{\textstyle\sum\limits_{i=1}^{k}}
\alpha(Z_{i})\Pi(d\varphi_{i},\beta)+\\
& +%
{\textstyle\sum\limits_{i,j=1}^{k}}
\alpha(Z_{i})\beta(Z_{j})\Pi(d\varphi_{i},d\varphi_{j}),
\end{align*}
so that the deformation $\Pi_{D}$ can be expressed as%
\begin{equation}
\Pi_{D}=\Pi-%
{\textstyle\sum\limits_{i}}
X_{i}\wedge Z_{i}+\frac{1}{2}%
{\textstyle\sum\limits_{i,j}}
\varphi_{ij}Z_{i}\wedge Z_{j}\label{pide}%
\end{equation}
where the functions $\varphi_{ij}$ are defined as%
\[
\varphi_{ij}=\left\{  \varphi_{i},\varphi_{j}\right\}  _{\Pi}=X_{j}%
(\varphi_{i})=\Pi(d\varphi_{i},d\varphi_{j}).
\]
In the Dirac case all the vector fields $X_{i}$ are transversal to the
foliation $\mathcal{S}$ and are moreover linearly independent. It happens
precisely when $\det(\varphi_{ij})\neq0$ (the functions $\varphi_{i}$ are then
'second class constraints' in the terminology of Dirac).The vector fields
$Z_{i}$ (the dual basis to $\left\{  d\varphi_{i}\right\}  $) can be expressed
through the vector fields $X_{i}$ as%
\[
Z_{i}=%
{\textstyle\sum\limits_{j=1}^{k}}
(\varphi^{-1})_{ji}X_{j}\text{, \ \ }i=1,\ldots,k
\]
Indeed,
\[
Z_{j}(\varphi_{i})=%
{\textstyle\sum\limits_{s=1}^{k}}
(\varphi^{-1})_{sj}X_{s}(\varphi_{i})=%
{\textstyle\sum\limits_{s=1}^{k}}
(\varphi^{-1})_{sj}\varphi_{is}=\delta_{ij}.
\]
Moreover, in this case the deformation (\ref{pide}) attains the form:%
\begin{equation}
\Pi_{D}=\Pi-\frac{1}{2}\sum_{i=1}^{k}X_{i}\wedge Z_{i}.\label{defdir}%
\end{equation}
This operator defines the following bracket on $C^{\infty}(\mathcal{M)}$
\begin{equation}
\{F,G\}_{\Pi_{D}}=\{F,G\}_{\Pi}-\sum_{i,j=1}^{k}\{F,\varphi_{i}\}_{\Pi
}(\varphi^{-1})_{ij}\{\varphi_{j},G\}_{\Pi},\label{Diracbracket}%
\end{equation}
which is just the well known \emph{Dirac deformation} \cite{Dirac} of the
bracket $\{.,.\}_{\Pi}$.

In the tangent case all the vector fields $X_{i}$ are tangent to the foliation
$\mathcal{S}$ \ and the deformation (\ref{pide}) attains the form%
\begin{equation}
\Pi_{D}=\Pi-%
{\textstyle\sum\limits_{i=1}^{k}}
X_{i}\wedge Z_{i},\label{deftan}%
\end{equation}
and has been considered in \cite{degiovanni} and in \cite{falquipedroni}.

In \cite{Diracrevisited} we have considered a bit more general then
(\ref{pide}) deformation of $\Pi$ of the form $\Pi_{d}=\Pi-\sum_{i}V_{i}\wedge
Z_{i}$. The image of $\Pi_{d}$ does not have to lie in $T\mathcal{S}$.
However, one can prove that it happens precisely when the vector fields
$V_{i}$ satisfy the following functional equation%
\begin{equation}
V_{i}=X_{i}+\sum_{j}V_{j}(\varphi_{i})Z_{j}.\label{func}%
\end{equation}
and in this case $\Pi_{d}$ is Poisson. Substituting this formula into $\Pi
_{d}$ yields
\begin{equation}
\Pi_{d}=\Pi-%
{\textstyle\sum\limits_{i=1}^{k}}
X_{i}\wedge Z_{i}+%
{\textstyle\sum\limits_{i,j=1}^{k}}
V_{j}(\varphi_{i})Z_{i}\wedge Z_{j},\label{pistare}%
\end{equation}
and it can be proved (after some technical manipulations) that in this case
$\Pi_{d}=\Pi_{D}$ (even though $V_{j}(\varphi_{i})$ does not have to be equal
to $\frac{1}{2}\varphi_{ij}$). A natural solution of the equation (\ref{func})
in the Dirac case is given by $V_{i}=\frac{1}{2}X_{i}$ and in the tangent case
by $V_{i}=X_{i}$, which turns (\ref{pistare}) into the deformations
(\ref{defdir}) and (\ref{deftan}) respectively. Thus, in a sense, the
deformation $\Pi_{D}$ is the canonical deformation in our setting.

Our construction is strongly related with the construction of Marsden and
Ratiu. J. Marsden and T. Ratiu presented in \cite{MarsdenRatiu} a natural way
of reducing of a given Poisson bracket $\left\{  \cdot,\cdot\right\}  _{\Pi}$
on $\mathcal{M}$ \ to a Poisson bracket $\left\{  \cdot,\cdot\right\}
_{\pi_{R}}$ on a given submanifold $\mathcal{S}_{\nu}$ (in our notation).
Their method is non-constructive in the sense that in order to find the
bracket $\left\{  f,g\right\}  _{\pi_{R}}$ of two functions $f$,$g:S_{\nu
}\rightarrow R$ one has to calculate $\mathcal{Z}|_{\mathcal{S}_{\nu}}%
$-invariant prolongations of these functions. Our construction is performed on
the level of bivectors rather than on the level of Poisson brackets. This
construction (by deformation of the bivector $\Pi$) applies directly to every
leaf of the distribution $\mathcal{S}$ \ and moreover it is constructive. At
every leaf, however, both constructions are equivalent. On the other hand, we
make the assumption about the transversality of the distribution $\mathcal{Z}$
\ that was not present in the original paper of Marsden and Ratiu. This
assumption is however very natural since it makes all the assumptions of
Poisson Reduction Theorem in \cite{MarsdenRatiu} automatically satisfied.

\section{Reduction of Hamiltonian dynamics}

Let us begin this section by stating - in our setting - a well-known theorem
about the relation between the Dirac deformation $\Pi_{D}$ of $\Pi$ and the
dynamic imposed by the constraints. Suppose, thus, that our manifold
$\mathcal{M}$ is a cotangent bundle of a Riemannian manifold $\mathcal{Q}$
with a covariant metric tensor $g$, so that $\mathcal{M}=T^{\ast}\mathcal{Q}
$. Denote the corresponding contravariant metric tensor by $G$. Considar a
Lagrangian dynamical system on $T\mathcal{Q}$
\begin{equation}
\frac{d}{dt}\frac{\partial L}{\partial\overset{\cdot}{q}_{i}}-\frac{\partial
L}{\partial q_{i}}=0,\ \ i=1,...,n,\label{lagr}%
\end{equation}
with a potential Lagrangian function $L(q,\overset{\cdot}{q})=\frac{1}%
{2}\overset{\cdot}{q}^{t}g\overset{\cdot}{q}-V(q)$. This of course leads to a
Hamiltonian equations of motion on $\mathcal{M}=T^{\ast}\mathcal{Q}$%
\begin{equation}
\overset{\cdot}{q}_{i}=\left\{  q_{i},H\right\}  _{\Pi}\text{ ,\ \ }%
\overset{\cdot}{p}_{i}=\left\{  p_{i},H\right\}  _{\Pi}\label{hamnonred}%
\end{equation}
with the Hamiltonian $H=\frac{1}{2}p^{t}Gp+V(q)$ and with the canonical
Poisson operator $\Pi$. Let us now impose a physical constraint $\varphi(q)=0$
on our system and assume that in the beginning the coordinates of the system
lie on the submanifold $\mathcal{Q}_{0}$ of $\mathcal{Q}$ given by
$\varphi(q)=0.$ One often makes a physical assumption here that the surface
$\mathcal{Q}_{0}$ starts to act on our system with a reaction force
$R(q,\overset{\cdot}{q})$ that is orthogonal to $\mathcal{Q}_{0}$ and such
that the trajectorties of the constrained system%
\begin{equation}
\frac{d}{dt}\frac{\partial L}{\partial\overset{\cdot}{q}_{i}}-\frac{\partial
L}{\partial q_{i}}=R_{i}(q,\overset{\cdot}{q})\label{zmodlagr}%
\end{equation}
that start on $\mathcal{Q}_{0}$ remain on $\mathcal{Q}_{0}$. On the level of
the phase space $\mathcal{M}=T^{\ast}\mathcal{Q}$ the Hamiltonian system
(\ref{hamnonred}) is now subordinate to a pair of constraints:%
\[
\varphi_{1}(q)\equiv\varphi(q)=0\text{, \ }\varphi_{2}(q)\equiv\left(
\nabla\varphi\right)  ^{t}Gp=0,
\]
where $\nabla\varphi$ is the gradient of $\varphi$ with respect to
$q$-variables (differential of $\varphi$ in $\mathcal{Q}$), (the second
constraint is a consequence of the fact that the velocities $\overset{\cdot
}{q}$ must remain tangent to $\mathcal{Q}_{0}$) and thus modifies to%
\begin{equation}
\overset{\cdot}{q}_{i}=\left\{  q_{i},H\right\}  _{\Pi}\text{ \ , \ }%
\overset{\cdot}{p}_{i}=\left\{  p_{i},H\right\}  _{\Pi}+R_{i}%
(q,p).\label{hamred}%
\end{equation}

\begin{theorem}
\bigskip The equations (\ref{hamred}) are Hamiltonian and can be written in
the form%
\begin{equation}
\overset{\cdot}{q}_{i}=\left\{  q_{i},H\right\}  _{\Pi_{D}}\text{ ,
\ \ }\overset{\cdot}{p}_{i}=\left\{  p_{i},H\right\}  _{\Pi_{D}}%
\label{zmodham}%
\end{equation}
where $\Pi_{D}=\Pi-\frac{1}{2}\sum_{i=1}^{2}X_{i}\wedge Z_{i}$ is the Dirac
deformation (\ref{defdir}) of $\Pi$ given by the constraints $\varphi_{1}$ and
$\varphi_{2}$.
\end{theorem}

Thus, the response of the Lagrangian system (\ref{lagr}) subordinated to the
reaction forces $R$ can be accounted for by the corresponding Direac
deformation of the Poisson operator $\Pi$. We will only sketch the proof.

\begin{proof}
The reaction force $R$ can be calculated by differentiating the assumed
identity $\varphi(q(t))\equiv0$ twice with respect to $t$ and eliminating the
second derivatives $\overset{\cdot\cdot}{q_{i}}$ with the help of equations
(\ref{zmodlagr}) and by using the demand that the force should be orthogonal
to $\mathcal{Q}_{0}$. After some calculations we obtain that:
\begin{equation}
R(q,p)=\frac{1}{\left(  \nabla\varphi\right)  ^{t}G\,\nabla\varphi}\left(
\left(  \nabla\varphi\right)  ^{t}G\,\nabla V-(p^{t}G)H_{\varphi
}(Gp)+A\right)  \nabla\varphi\label{Rwprost}%
\end{equation}
where $H_{\varphi}$ is the Hessian of $\varphi$: $\left(  H_{\varphi}\right)
_{ij}=\frac{\partial^{2}\varphi}{\partial q_{i}\partial q_{j}}$ and where
$A=A(q,p)$ is given by%
\[
A=%
{\textstyle\sum\limits_{s}}
\frac{\partial\varphi}{\partial q_{s}}%
{\textstyle\sum\limits_{r,m,i,j}}
\Gamma_{ij}^{s}G^{ir}G^{jm}p_{r}p_{m}%
\]
so that it vanishes in the Euclidean coordinates when all Christoffel's
symbols $\Gamma_{jk}^{i}$ are equal to zero. On the other hand, calculating
the explicit form of (\ref{zmodham}) on the submanifold of $\mathcal{M}%
=T^{\ast}\mathcal{Q}$ given by the constraints $\varphi_{1},\varphi_{2}$ leads
to the equations (\ref{hamred}) with $R$ given by (\ref{Rwprost}).
\end{proof}

Let us now consider a Hamiltonian vector field $X=\Pi dH$ on a general Poisson
manifold $\mathcal{M}$ where $H$ is some real-valued smooth function on
$\mathcal{M}$ (Hamiltonian function). We constantly assume that we have a
smooth, regular foliation $\mathcal{S}$ on $\mathcal{M}$ and a regular
distribution $\mathcal{Z}$ on $\mathcal{M}$ such that (\ref{split}) is
satisfied. The corresponding $\Pi_{D}$ defined by (\ref{defpi}) is Poisson and
has its image tangent to the foliation $\mathcal{S}$, so that it can be
properly restricted on every leaf $\mathcal{S}_{\nu}$ of $\mathcal{S}$. Thus,
the following definition makes sense.

\begin{definition}
We call the vector field $X_{D}=\Pi_{D}dH$ the Hamiltonian projection of the
Hamiltonian vector field $X=\Pi dH$.
\end{definition}

The vector field $X_{D}$ lives on every leaf of the foliation in the sense
that its restriction on the leaf $\mathcal{S}_{\nu}$ is tangent to
$\mathcal{S}_{\nu}$. Moreover, on the leaf $\mathcal{S}_{\nu}$ it coincides
with the Hamiltonian vector field $\pi_{R_{\nu}}dh$:
\[
\left.  \Pi_{D}dH\right\vert _{\mathcal{S}_{\nu}}=\pi_{R_{\nu}}dh,
\]
where $h=\left.  H\right\vert _{\mathcal{S}_{\nu}}$ is the restriction of the
Hamiltonian $H$ to the leaf $\mathcal{S}_{\nu}$. To see this it is enough to
choose a parametrization $\left\{  \varphi_{i}\right\}  $ of $\mathcal{S}$ and
to pass to any system of coordinates of the form $(x,\varphi)$. In these
coordinates the bivector $\Pi_{D}$ has a matrix form with a non-zero
upper-left block coinciding with the matrix form of $\pi_{R_{\nu}}$ and with
the remaining terms equal to zero.

There is a connection between $X=\Pi dH$ and its Hamiltonian projection
$X_{D}=\Pi_{D}dH$.

\begin{theorem}
\label{dynamic}(Dynamic reduction theorem) If $X=\Pi dH$, $X_{D}=\Pi_{D}dH$
and $X_{i}=\Pi d\varphi_{i}$ then
\begin{equation}
X_{D}=X_{\parallel}-%
{\textstyle\sum\limits_{i=1}^{k}}
Z_{i}(H)X_{i,\parallel}.\label{difference}%
\end{equation}

\end{theorem}

\begin{proof}
A direct calculation yields%
\begin{align*}
X_{D}  & =\Pi_{D}dH=X-%
{\textstyle\sum\limits_{i=1}^{k}}
\left(  Z_{i}(H)X_{i}-X_{i}(H)Z_{i}\right)  +%
{\textstyle\sum\limits_{i,j=1}^{k}}
\varphi_{ij}Z_{j}(H)Z_{i}=\\
& =X_{\parallel}-%
{\textstyle\sum\limits_{i=1}^{k}}
Z_{i}(H)\left(  X_{i}-%
{\textstyle\sum\limits_{j=1}^{k}}
\varphi_{ji}Z_{j}\right)  ,
\end{align*}
the last equality due to $X_{i}(H)=\left\langle dH,\Pi d\varphi_{i}%
\right\rangle =-\left\langle d\varphi_{i},\Pi dH\right\rangle =-X(\varphi
_{i})$. Since $\varphi_{ji}=X_{i}(\varphi_{j})$ it yields (\ref{difference}).
\end{proof}

Observe that the difference between $X_{D}$ and $X_{\parallel}$ is the term $%
{\textstyle\sum\limits_{i=1}^{k}}
Z_{i}(H)X_{i,\parallel}$ that is tangent to the foliation $\mathcal{S}$, as it
should be.

Since for the Dirac case $X_{i,\parallel}=0$ (by definition) we have

\begin{corollary}
In the Dirac case $X_{D}=X_{\parallel}$, so that in this case the Hamiltonian
projection is just the natural projection (in the sense of direct sum) along
the distribution $\mathcal{Z}$.
\end{corollary}

The term $X_{\parallel}$ in $X_{D}$ has a well-known physical interpretation:
it describes the evolution of the system $X=\Pi dH$ imposed with the
constraints given by $\varphi_{i}$. The physical meaning of the second term in
$X_{D}$ is not clear for us, although it should represent some additional
force (friction) acting on the system and tangent to the constraints. The
authors are grateful for any hints in this matter.

We are now in position to discuss the degeneracy of $\Pi_{D}$ using the above
Dynamic reduction theorem (Theorem \ref{dynamic}). Let us first discuss the
Dirac case.

\begin{proposition}
Consider the Dirac deformation $\Pi_{D}$ given by (\ref{defdir}) of a Poisson
operator $\Pi$ on M. Suppose that the real valued functions $c_{i}$,
$i=1,\ldots,s$ on $\mathcal{M}$ are such that they span the kernel of the
operator $\Pi$ (i.e. are Casimir functions of $\Pi$ in the sense that $\Pi
dc_{i}=0$) and such that the functions $\left\{  c_{i},\varphi_{j}\right\}  $
constitute a functionally independent set. Then

i) the constraints $\varphi_{i}$ and the 'old' Casimirs $c_{i}$ are all
Casimirs of $\Pi_{D}$

ii) any Casimir of $\Pi_{D}$ must be of the form $C(c_{1},\ldots,c_{s}%
,\varphi_{1},\ldots\varphi_{k})$.
\end{proposition}

\begin{proof}
The proof of i) is just a calculation. To prove ii), let us complete the
functionally independent functions $\left\{  c_{i},\varphi_{j}\right\}  $ to a
coordinate system $\left\{  c_{i},\varphi_{j},x_{k}\right\}  $ on
$\mathcal{M}$. Suppose that a function $C=C(c,\varphi,x)$ is a Casimir of
$\Pi_{D}$ given by (\ref{defdir}), i.e. that $\Pi_{D}dC=0$. Then, according to
Theorem \ref{dynamic}, $\left(  \Pi dC\right)  _{\parallel}=0$, i.e. $\Pi
dC\subset\mathcal{Z}$. In the Dirac case the distribution $\mathcal{Z}$ is
spanned by the vector fields $X_{i}$ so that there must exist functions
$\alpha_{i}$ such that $\Pi dC=%
{\textstyle\sum\nolimits_{i=1}^{k}}
\alpha_{i}X_{i}=%
{\textstyle\sum\nolimits_{i=1}^{k}}
\alpha_{i}\Pi d\varphi_{i}=\Pi\left(
{\textstyle\sum\nolimits_{i=1}^{k}}
\alpha_{i}d\varphi_{i}\right)  $. Thus, $\Pi\left(  dC-%
{\textstyle\sum\nolimits_{i=1}^{k}}
\alpha_{i}d\varphi_{i}\right)  =0$ or $dC=%
{\textstyle\sum\nolimits_{i=1}^{k}}
\alpha_{i}d\varphi_{i}+%
{\textstyle\sum\nolimits_{i=1}^{s}}
\beta_{i}dc_{i}$ which proves ii).
\end{proof}

Thus, we can state that the Dirac deformation (\ref{defdir}) preserves all the
old Casimir functions and introduces new Casimirs $\varphi_{i}$ and that no
other Casimirs arise in this process. The situation in the general case is
more complicated, since the Casimirs of $\Pi$ does not have to survive the
general deformation (\ref{defpi}) (or (\ref{pide})) and moreover since new
Casimirs, different from $\varphi_{i}$ ones, can arise. We can merely state
that in the general case the function $C$ is a Casimir of $\Pi_{D}$ if and
only if the vector field $Y=\Pi dC$ satisfies the relation%
\[
Y_{\parallel}=%
{\textstyle\sum\limits_{i=1}^{k}}
Z_{i}(C)X_{i,\parallel}%
\]
which for the Dirac case degenerates to the already discussed condition
$Y_{\parallel}=0$.

\section{Example}

Let us conclude this article by an example. Consider the so called first
Newton representation of the seventh-order stationary flow of the KdV
hierarchy \cite{myPhysicaA},\cite{1},\cite{Diracpencils}. It is the following
system of second order Newton equations
\begin{align*}
q_{1,tt}  & =-10q_{1}{}^{2}+4q_{2}\\
q_{2,tt}  & =-16q_{1}q_{2}+10q_{1}{}^{3}+4q_{3}\\
q_{3,tt}  & =-20q_{1}q_{3}-8q_{2}{}^{2}+30q_{1}^{2}q_{2}-15q_{1}^{4}%
\end{align*}
(where the subscript $,t$ denotes the differentiation with respect to the
evolution parameter $t$). By putting $p_{1}=q_{3,t}$, $p_{2}=q_{2,t}$,
$p_{3}=q_{1,t}$ it can be written in a Hamiltonian form:%
\[
\frac{d}{dt}\left(  q_{1},q_{2},q_{3},p_{1},p_{2},p_{3}\right)  ^{T}=X=\Pi dH
\]
where $\Pi$ is the canonical Poisson operator on the space $\mathcal{M}%
=\left\{  \left(  q_{1},q_{2},q_{3},p_{1},p_{2},p_{3}\right)  \right\}  $,%
\[
\Pi=\left[
\begin{array}
[c]{c|c}%
0 & I_{3}\\\hline
-I_{3} & 0
\end{array}
\right]  \text{ \ }%
\]
and with the Hamiltonian%

\[
H=p_{1}p_{3}+\tfrac{1}{2}p_{2}^{2}+10q_{1}{}^{2}q_{3}-4q_{2}q_{3}+8q_{1}%
q_{2}{}^{2}-10q_{1}{}^{3}q_{2}+3q_{1}{}^{5}.
\]
Consider also a foliation $\mathcal{S}$ given by a pair of constraints%
\[
\varphi_{1}=q_{3}+q_{1}q_{2}\text{, \ \ \ }\varphi_{2}=p_{1}+q_{1}p_{2}%
+q_{2}p_{3}\text{.}%
\]
where $\varphi_{2}$ is the so called $G$-consequence of $\varphi_{1}$ (i.e. a
lift from the configuration space $\left\{  \left(  q_{1},q_{2},q_{3}\right)
\right\}  $ to $\mathcal{M}$) with respect to the antidiagonal metric tensor
$G$ \cite{Diracpencils}. The vector fields $X_{i}=\Pi d\varphi_{i}$ have the
form%
\[
X_{1}=-q_{2}\frac{\partial}{\partial p_{1}}-q_{1}\frac{\partial}{\partial
p_{2}}-\frac{\partial}{\partial p_{3}}\text{, \ }X_{2}=\frac{\partial
}{\partial q_{1}}+q_{1}\frac{\partial}{\partial q_{2}}+q_{2}\frac{\partial
}{\partial q_{3}}-p_{2}\frac{\partial}{\partial p_{1}}-p_{3}\frac{\partial
}{\partial p_{2}}%
\]
and they are transversal to $\mathcal{S}$ so that we have the Dirac case.
Thus, the distribution $\mathcal{Z}=\mathcal{D}=Sp\left\{  X_{i}\right\}  $
makes $\Pi$ $\mathcal{Z}$-invariant. The basis in $\mathcal{Z}$ that is dual
to $\left\{  d\varphi_{i}\right\}  $ is $Z_{1}=\frac{1}{\varphi_{12}}X_{2}$,
$Z_{2}=-$ $\frac{1}{\varphi_{12}}X_{1}$, where $\varphi_{12}=\left\{
\varphi_{1},\varphi_{2}\right\}  _{\Pi}=2q_{2}+q_{1}^{2}$. The Dirac
deformation $\Pi_{D}$ given by (\ref{defdir}) attains in the adapted
coordinate system $\left\{  \left(  q_{1},q_{2},\varphi_{1},\varphi_{2}%
,p_{2},p_{3}\right)  \right\}  $ the form%
\[
\Pi_{D}=\frac{1}{2q_{2}+q_{1}^{2}}\left[
\begin{array}
[c]{cccccc}%
0 & 0 & 0 & 0 & -q_{1} & -1\\
0 & 0 & 0 & 0 & 2q_{2} & -q_{1}\\
0 & 0 & 0 & 0 & 0 & 0\\
0 & 0 & 0 & 0 & 0 & 0\\
q_{1} & -2q_{2} & 0 & 0 & 0 & p_{3}\\
1 & q_{1} & 0 & 0 & -p_{3} & 0
\end{array}
\right]
\]
It has, as it should, two Casimirs $\varphi_{1},\varphi_{2}$. We can now
easily restrict $\Pi_{D}$ to the operator $\pi_{R_{\nu}}$ on $\mathcal{S}%
_{\nu}$. If we parametrize $\mathcal{S}_{\nu}$ with the coordinates $\left\{
\left(  q_{1},q_{2},p_{2},p_{3}\right)  \right\}  $ (the constraints
$\varphi_{1},\varphi_{2}$ are constant on every $\mathcal{S}_{\nu}$) then%
\[
\pi_{R_{\nu}}=\frac{1}{2q_{2}+q_{1}^{2}}\left[
\begin{array}
[c]{cccc}%
0 & 0 & -q_{1} & -1\\
0 & 0 & 2q_{2} & -q_{1}\\
q_{1} & -2q_{2} & 0 & p_{3}\\
1 & q_{1} & -p_{3} & 0
\end{array}
\right]  ,
\]
which, in accordance with the theory, is non-degenerate. Observe that this
expression actually does not depend on the choice of the leaf $\mathcal{S}%
_{\nu}$ in the foliation $\mathcal{S}$. The Hamiltonian projection $X_{D}%
=\Pi_{D}dH$ attains on every leaf $\mathcal{S}_{\nu}$ the form%
\[
\pi_{R_{\nu}}dh_{R_{\nu}}=\frac{1}{2q_{2}+q_{1}^{2}}\left(  \alpha_{1}%
\frac{\partial}{\partial q_{1}}+\alpha_{2}\frac{\partial}{\partial q_{2}%
}+\beta_{2}\frac{\partial}{\partial p_{2}}+\beta_{3}\frac{\partial}{\partial
p_{3}}\right)
\]
where the functions $\alpha_{i}$, $\beta_{i}$ are some rather complicated
polynomial functions of coordinates and the parameters $\nu_{i}$ of the leaf
$\mathcal{S}_{\nu}$.

Let us now choose another distribution $\mathcal{Z}$, for which $\Pi$ is also
$\mathcal{Z}$-invariant. Since the operator $\Pi$ has a very simple form, any
pair of constant fields will span a distribution $\mathcal{Z}$ that makes
$\Pi$ -invariant, since then the Vaisman condition (\ref{vais}) is trivially
satisfied. Thus, let us take $\mathcal{Z}=Sp\left\{  \frac{\partial}{\partial
q_{3}}+\frac{\partial}{\partial p_{3}},\frac{\partial}{\partial q_{3}}%
+\frac{\partial}{\partial p_{1}}\right\}  $ (observe that this distribution is
integrable). We have now to change the basis in $\mathcal{Z}$ to a new basis
$\left\{  Z_{1},Z_{2}\right\}  $ such that the condition $Z_{i}(\varphi
_{j})=\delta_{ij}$ is satisfied. A simple calculation yields%

\[
Z_{1}=\frac{\partial}{\partial q_{3}}+\frac{1}{1-q_{2}}\left(  -q_{2}%
\frac{\partial}{\partial p_{1}}+\frac{\partial}{\partial p_{3}}\right)
\text{, \ }Z_{2}=\frac{1}{1-q_{2}}\left(  \frac{\partial}{\partial p_{1}%
}-\frac{\partial}{\partial p_{3}}\right)  .
\]
Now, the general deformation (\ref{pide}) defined by the above distribution
attains in the coordinates $\left\{  \left(  q_{1},q_{2},\varphi_{1}%
,\varphi_{2},p_{2},p_{3}\right)  \right\}  $ the form
\[
\Pi_{D}=\frac{1}{q_{2}-1}\left[
\begin{array}
[c]{cccccc}%
0 & 0 & 0 & 0 & 0 & -1\\
0 & 0 & 0 & 0 & q_{2}-1 & -q_{1}\\
0 & 0 & 0 & 0 & 0 & 0\\
0 & 0 & 0 & 0 & 0 & 0\\
0 & 1-q_{2} & 0 & 0 & 0 & p_{3}-q_{1}\\
1 & q_{1} & 0 & 0 & q_{1}-p_{3} & 0
\end{array}
\right]
\]
and thus the restricted operator $\pi_{R_{\nu}}$ on the leaf $\mathcal{S}%
_{\nu}$ parametrized with the coordinates $\left\{  \left(  q_{1},q_{2}%
,p_{2},p_{3}\right)  \right\}  $ is%

\[
\pi_{R_{\nu}}=\frac{1}{q_{2}-1}\left[
\begin{array}
[c]{cccc}%
0 & 0 & 0 & -1\\
0 & 0 & q_{2}-1 & -q_{1}\\
0 & 1-q_{2} & 0 & p_{3}-q_{1}\\
1 & q_{1} & q_{1}-p_{3} & 0
\end{array}
\right]  ,
\]
which is again non-degenerate. Again, this expression does not depend on the
choice of the leaf $\mathcal{S}_{\nu}$ in the foliation $\mathcal{S}$.

In the end, consider yet another distribution that makes $\Pi$ $\mathcal{Z}%
$-invariant, namely $\mathcal{Z}=Sp\left\{  \frac{\partial}{\partial q_{3}%
}+\frac{\partial}{\partial p_{3}},\frac{\partial}{\partial p_{1}}\right\}  $.
The appropriate basis of $\mathcal{Z}$ is given by%

\[
Z_{1}=\frac{\partial}{\partial q_{3}}-q_{2}\frac{\partial}{\partial p_{1}%
}+\frac{\partial}{\partial p_{3}}\text{, \ }Z_{2}=\frac{\partial}{\partial
p_{1}}.
\]
so that we have $Z_{i}(\varphi_{j})=\delta_{ij}$. The general deformation
(\ref{pide}) defined by the above distribution yields in the coordinates
$\left\{  \left(  q_{1},q_{2},\varphi_{1},\varphi_{2},p_{2},p_{3}\right)
\right\}  $
\[
\Pi_{D}=\frac{1}{q_{2}-1}\left[
\begin{array}
[c]{cccccc}%
0 & 0 & 0 & 0 & 0 & 0\\
0 & 0 & 0 & 0 & 1 & 0\\
0 & 0 & 0 & 0 & 0 & 0\\
0 & 0 & 0 & 0 & 0 & 0\\
0 & -1 & 0 & 0 & 0 & q_{1}\\
0 & 0 & 0 & 0 & -q_{1} & 0
\end{array}
\right]
\]
so that the restricted operator $\pi_{R_{\nu}}$ on the leaf $\mathcal{S}_{\nu
}$ (again parametrized with the coordinates $\left\{  \left(  q_{1}%
,q_{2},p_{2},p_{3}\right)  \right\}  $) is%

\[
\pi_{R_{\nu}}=\frac{1}{q_{2}-1}\left[
\begin{array}
[c]{cccc}%
0 & 0 & 0 & 0\\
0 & 0 & 1 & 0\\
0 & -1 & 0 & q_{1}\\
0 & 0 & -q_{1} & 0
\end{array}
\right]  ,
\]
and is degenerate this time.

Thus, given a foliation $\mathcal{S}$, by choosing different distributions
$\mathcal{Z}$ we can obtain several different Hamiltonian projections of our
original Hamiltonian system, not only just the well known Dirac reduction.

\section{Conclusions}

In this article we formulated a comprehensive, geometric picture of what is
know as Dirac reduction and Marsden-Ratiu reduction of a Poisson operator
$\Pi$ on a foliation $\mathcal{S}$ not related with the symplectic foliation
of $\Pi$. As a consequence, we obtain a geometric method of reducing of any
Hamiltonian system to a Hamiltonian system on the foliation $\mathcal{S}$. Any
such reduction depends merely on the choice of the distribution $\mathcal{Z}$
along which the reduction take place. Thus, we can reduce Hamiltonian systems
on $\mathcal{M}$ to Hamiltonian systems on $\mathcal{S}$ in many
non-equivalent ways. However, the procedure of finding appropriate
distributions $\mathcal{Z}$ (i.e. those that make $\Pi$ $\mathcal{Z}%
$-invariant) is non-trivial and non-algorithmic.


\begin{thebibliography}{99}                                                                                               %
\bibitem {Van der Schaft}A. J. van der Schaft, B. M. Maschke, "On the
Hamiltonian formulation of nonholonomic mechanical systems",
$\emph{Rep.Math.Phys.}$ \textbf{34} (1994), no. 2, 225--233.

\bibitem {Dirac}P. A. M. Dirac, \textquotedblright Generalized Hamiltonian
Dynamics\textquotedblright, \emph{Can. J. Math}. \textbf{2} (1950) 129--148.

\bibitem {blan}G. Blankenstein and A. J. van der Schaft, \emph{Symmetry and
reduction in implicit generalized Hamiltonian systems, }Rep. Math. Phys.
\textbf{47 }(2001) 57-100

\bibitem {s1}J. \'{S}niatycki, \emph{Dirac brackets in geometric dynamics,
}Ann. Inst. H. Poincar\'{e} Sect. A (N.S.) \textbf{20 }(1974) 365-372

\bibitem {lichnerowicz}A. Lichnerowicz, $\emph{Vari}\acute{e}\emph{t}\acute
{e}$ $\emph{symplectique}$ $\emph{et}$ $\emph{dynamique}$ $\emph{associ}%
\acute{e}\emph{e}$ $\grave{a}$ $\emph{une}$ $\emph{sous-vari}\acute{e}%
\emph{t}\acute{e}$, C.R.Acad.Sci.Paris Ser. A-B \textbf{280} (1975), A523-A527

\bibitem {Lichnerowicz}A. Lichnerowicz, Les vari\'{e}t\'{e}s de Poisson et
leurs alg\`{e}bres de Lie associ\'{e}es, \emph{J. Diff. Geom.} \textbf{12}
(1977), 253--300.

\bibitem {mar1}J. E. Marsden and A. Weinstein, \emph{Reduction of symplectic
manifolds with symmetry, }Rep. Math. Phys. \textbf{5 }(1974) 121-130

\bibitem {MarsdenRatiu}Marsden and T. Ratiu, \emph{Reduction of Poisson
manifolds, }Lett. Math. Phys. \textbf{11 }(1986) 161-169

\bibitem {Marmo}J. Grabowski, G. Landi, G. Marmo and G. Vilasi,
\emph{Generalized Reduction Procedure: Symplectic and Poisson Formalism}
Fortschr. Phys. \textbf{42} (1994) 393-427

\bibitem {fla}M. Flato, A. Lichnerowicz and D. Sternheimer, \emph{Deformations
of Poisson brackets, Dirac brackets and applications, }J. Math. Phys.
\textbf{17 }(1976) 1754--1762

\bibitem {Diracrevisited}K. Marciniak, M. Blaszak, "Dirac reduction
revisited", \emph{J. Nonlin. Math. Phys.} \textbf{10} (2003) 451--463, or arXiv.org/nlin.SI/0303014.

\bibitem {degiovanni}L. Degiovanni, G. Magnano, "Tri-Hamiltonian vector
fields, spectral curves and separation coordinates", \emph{Rev. Math. Phys.}
\textbf{14} (2002) 115--1163.

\bibitem {falquipedroni}G. Falqui, M. Pedroni, "Separation of variables for
Bi-Hamiltonian systems", $\emph{Math.Phys.Anal.Geom}.$ \textbf{6} (2003) 139--179.

\bibitem {Marle}Ch.-M. Marle, "Various approaches to conservative and
nonconservative nonholonomic systems", \emph{Rep. Math. Phys.} \textbf{42}
(1998) 211--229.

\bibitem {dualpp}M. Blaszak, K. Marciniak, "Dirac reduction of dual
Poisson-presymplectic pairs", \emph{J. Phys. A: Math. Gen. } \textbf{37}
(2004) 5173--5187.

\bibitem {Vaisman}I. Vaisman, "Lectures on the Geometry of Poisson Manifolds".
Progress in Math., Birkh\"{a}user 1994.

\bibitem {Diracpencils}M. Blaszak, K. Marciniak, "Separability preserving
Dirac reductions of Poisson pencils on Riemannian manifolds", \emph{J. Phys.
A: Math. Gen.} \textbf{36} (2003), 1337--1356.\emph{\ }

\bibitem {myPhysicaA}S. Rauch-Wojciechowski, K. Marciniak, M. Blaszak,
\textquotedblright Two Newton decompositions of stationary flows of KdV and
Harry Dym hierarchies\textquotedblright\ \emph{Physica A} \textbf{233} (1996) pp.307--330

\bibitem {1}B\l aszak M, \textquotedblright On separability of bi-Hamiltonian
chain with degenerated Poisson structures\textquotedblright, \emph{J. Math.
Phys.} \textbf{39, }3213 (1998)
\end{thebibliography}
\end{document}